\documentclass[a4paper]{article}
\begin{document}

\title{Evidence for radiative generation of lepton masses}
\author{Hans de Vries\thanks{ {\tt hansdevries@chip-architect.com}}, 
Alejandro Rivero\thanks{Zaragoza University at Teruel.  
           {\tt arivero@unizar.es}}}

\maketitle

\begin{abstract}
       We present a fit to the experimental charged lepton masses
as coming from radiative corrections in QED.
\end{abstract}

\section{Introduction}

In November of the last year, one of us (HdV) noticed that the values
of $a_e \equiv (g_e-2)/2$, the anomalous moment of electron, and of the difference
$a_\mu-a_e$ with the one of the muon, were amazingly close to the
mass quotients $m_\mu/m_Z$ and $m_e/m_W$. This happened during an
on-going internet quest for accurate empirical relationships between 
fundamental constants, but we felt that the accuracy of this particular
case deserved further investigation:

\begin{quote}
----------------------------------------------------------------------------------------
\end{quote}
$$
  \begin{array}{lll}
  0.00115869 & = & \mbox{muon / Z mass ratio} \\
  0.00115965 & = & \mbox{electron magnetic anomaly} \\
  0.00000635 & = & \mbox{electron / W mass ratio} \\
  0.00000626 & = & \mbox{difference of muon and electron magnetic
  anomaly}
\end{array}
$$
\begin{quote}
-------------------------------------- table 1.-----------------------------------------
\end{quote}


Of course the calculation of $(g-2)/2$ involves the very well known series
on the electromagnetic couping $\alpha$. A coincidence with simple
combinations of lepton masses can be explained if such masses 
come themselves from expressions containing $\alpha$. Then it 
strongly suggests that such masses are generated radiatively
in such way that at low order both perturbative
series can be related. It has been observed from time ago \cite{barr} 
that lepton masses have quotients of order $\alpha$, and a whole
industry of model-making starts from trying to fit it \cite{barr,
giorgi}, getting the masses as radiative series on $\alpha$. But until
now, no new evidence had been observed for this kind of schemes

\section{Self Energy and Vacuum Polarization}
As we expect only a parallel between structures, we can do the ansatz
of comparing the first quotient exclusively to self-energy graphs, and to
ascribe all of the vacuum polarisation (v.p.) contribution 
to the second. This ansatz was
guided by Hans' observation of the similarity between the quotient
$m_W/m_Z$ and the ratio of semiclassical\footnote{$\beta_s$ is the velocity of a 
mass rotating on a orbit with angular momentum $\sqrt{s(s+1)} \hbar$ and
a frequency corresponding to its rest mass. The quotient $\beta_\frac12/\beta_1$
is about 0.8814}
velocities $\beta_1/\beta_\frac12$. In any case, it amounts to excluding
the electron vacuum polarisation loop in the $\alpha^2$ order and, because
precision requires it, the v.p. and light by light diagrams in 
the third order. These are, respectively \cite{lw}  

\begin{equation}
a_e^{vp}= ({119\over36}-\frac13 \pi^2) (\frac\alpha\pi)^2 
-0.099 (\frac\alpha\pi)^3 + 0.37 (\frac\alpha\pi)^3= 88.0\ 10^{-9}
\end{equation}

Our table becomes

\begin{quote}
----------------------------------------------------------------------------------------
\end{quote}
\begin{quote}
\begin{tabular}{lcl}
0.001 158 692 3 &=& muon / Z mass ratio \\
0.001 159 564 2 &=& exp. $a_e$ electron magnetic anomaly - $a_e^{vp}$\\
0.000 000 871 9 &=& difference \\
                & & \\
0.001 165 046 0 &=& muon / Z mass ratio + electron / W mass ratio     \\
0.001 165 920 8 &=& exp muon magnetic anomaly $a_\mu$    \\
0.000 000 874 8 &=& difference \\
                & & \\
0.000 006 353 7 &=& electron / W mass ratio \\    
0.000 006 356 7 &=& exp. $a_\mu- \mbox{exp. }a_e + a_e^{vp}) $ \\
0.000 000 003 0 &=& difference \\
\end{tabular}
\end{quote}

\begin{quote}
-------------------------------------- table 2.-----------------------------------------
\end{quote}

The uncertainty due to W mass is $2.99\cdot 10^{-9}$. Actually, the third loop perturbation,
above incorporated, has a positive contribution $3.4\cdot 10^{-9}$ against a perfect match. In
any case, it is empirically very satisfying to find oneself inside the experimental
error with only an ansatz on the diagrams. Still, the $\mu/Z$ ratio is accurate only up to
$O(\frac\alpha\pi)$, and it seems to ask for an additional $O(({\alpha\over\pi})^2)$ term.

\section{First QED approximation}

So, lets try this ansatz in a pure calculational setup, without recurring to the
experimental data, and lets see if -or how- the coincidence can be related to a
parallell of mathematical structures. The QED calculation of $a_e$ excluding
vacuum polarisation 
is \cite{lw},
\begin{equation}a_e^{QED-v.p}
=\frac12 {\alpha \over \pi} - 0.344 166 8 ({\alpha \over \pi})^2
 +0.943 ({\alpha \over \pi})^3
=0.001\ 159\ 564\ 60
\end{equation}
while the whole QED result for $a_\mu$ is\cite{cm}
\begin{equation}
a_\mu^{QED}=0.5 {\alpha \over \pi} + 0.765 857 388 ({\alpha
 \over \pi})^2 + 24.0505 ({\alpha \over \pi})^3 + 126.04 ({\alpha \over \pi})^4
 =0.001\ 165\ 847\ 00 
\end{equation}

The difference being $a_\mu-a_e=0.000 006 282 40$
The coincidences are thus initially of 99.92\% and 98.88\% and 
by themselves they should constitute at least collateral evidence of radiative
terms for most part of the $m_e, m_\mu$. Note that by betting for a 
mathematical structure with leptons only, we have lost the hadronic 
contribution, of order $67\cdot 10^{-9}$, so now we are too 
in need of a corrective term 
for the missing 01.12\% if we want to increase the order of accuracy.

\section{Additional Terms}

Our first research must be how the fit to $m_\mu / m_Z$ can be improved 
by using additional terms.
There is no very much playroom using only electroweak mass data, but 
a bit surprisingly, there are possibilities of improvement. Keeping with simple 
quotients of Z, it is possible to enter into the one-sigma experimental
precision of Z, $26.68\cdot 10^{-9}$, by using $(1 / 2 \pi) m_e / m_Z$. If we are willing 
to admit more higher powers of mass quotients, a term $m_\mu^2/2 m_W^2$ 
drives the estimate up to almost full coincidence with the central values.
And if we do not like extra coefficients, we can instead use $m_\mu^2/ m_X^2$
for an undiscovered mass X of 114.5 GeV. Let us compare these possibilities:

\begin{eqnarray}
\label{nosquare}
a_e^{QED-v.p} &=0.001 159 564 60 \\
{m_\mu \over m_Z} +{1 \over 2 \pi} {m_e \over m_Z}&=0.001 159 584 17  &:-0.000\ 000\ 019\ 57 \\  
\label{wcorr}
{m_\mu \over m_Z} +{m_\mu^2 \over 2 m_W^2}&=0.001 159 555 26 &: 0.000\ 000\ 009\ 34 \\
{m_\mu \over m_Z}+{m_\mu^2 \over m_X^2}&=.001 159 543 81 &: 0.000\ 000\ 020\ 79 \\
               & \mbox {Z error}  &: 0.000\ 000\ 026\ 68
\end{eqnarray}

The last column shows differences, to be compared with the uncertainness induced from the 
experimental measurement of $Z_0$.

The fit at X is appealing because Z is a neutral particle, and the experimental
hint of CERN at this value was for the neutral scalar. While waiting for news in the
experimental front, we can happily admit the correction of (\ref{wcorr}). 

Another motivation to prefer quadratic correction terms is that we can use also a term
in $m_e m_\tau$ to recover almost completely the precision we lost for the second quotient when
we decided to do not include the hadronic (quark loop) contributions. We have 
\begin{eqnarray}
a_\mu^{QED}-a_e^{QED-v.p}&=0.000 006 282 40 \\
\label{taucorr}
{m_e \over m_W} - {m_e m_\tau \over 2 m_W^2}&=0.000 006 283 54 &: 0.000\ 000\ 001\ 14 \\
{m_e \over m_W} - {m_e m_\tau \over m_X^2}&=0.000 006 284 47  &:0.000\ 000\ 002\ 07  \\
& \mbox {W error} &: 0.000\ 000\ 002\ 99
\end{eqnarray}
Again, the last column shows differences, to be compared with the uncertainness induced from the 
experimental measurement of $W^+$. And besides the already mentioned hadronic 
contribution, $67\cdot 10^{-9}$, we could  consider also the pure electroweak 
contribution, $1.51\cdot 10^{-9}$, to be added to $a_\mu$. We mention it
separately  to show that we can not decide if we are comparing against 
the structure of a pure QED kind of series or against an electroweak series.

\section{Remarks}

{\bf Remark 1}. 
It can be asked if there is a role for the tau anomalous moment in this scheme.
It is a touchy issue, because while the tau lives at order $\alpha$ of the 
electroweak vacuum\footnote{As Jay R Yablon reminded us recently}, it is also
at the mass scale typical of SU(3) colour, while the next lepton, the muon,
lives at the mass scale of the chiral breaking (whose goldstone boson is
in some sense the pion). We can suspect things are not very clear cut in
its radiative process, and in fact one could prefer to admit quarks in
the calculation instead of using the correction of 
formula (\ref{taucorr}) above, and then to adjust the $a_e$ term with 
formula (\ref{nosquare}).

As for the $a_\tau$ correction it refers,  it is tempting to try to guess 
if a simple expression does it exist. This value is not
known experimentally, but from Samuel et al \cite{S}, we know its calculated
QED value, 0.0011732. If we ask for a simple quotient,
we would again to use the electron mass over some particle $X^+$, which
we could expect (but not necessarily) to be a charged one, to 
imitate the use of W. The total expression
\begin{equation}
{m_\mu\over m_Z} + {m_e\over m_W} + {m_e\over m_{X^+}} +
{m_\mu^2- m_e m_\tau \over m_X^2}
\end{equation}
actually matches $a_\tau$ for a mass of $X^+$ about 68 GeV \footnote{It 
is perhaps
worth to note here that the existence of a charged scalar at this
value was pursued \cite{L3} in the LEP, while the final evaluation
reduced the value of the events down to a two sigma deviation. So in
some sense this scalar has presently the same experimental status
that the events at 115 GeV assigned to a neutral scalar.}. 

{\bf Remark 2}.
In principle, if all the three formulae above are taken seriously, a
matching order-by-order in $\alpha$ could be done to estimate
the corresponding coefficients of the radiative series for each lepton
mass. But without further understanding of the role of the
electroweak bosons, or of the full electroweak scale and the role
of $\tau$, such matching becomes merely a mathematical exercise.

{\bf Remark 2.5}
Another consequence of taking seriously the quadratic formulae is that 
their simultaneous use gives an hyperbolic relationship between 
electroweak masses. It should be interesting if some family of GUT models
were able to generate this kind of relationship:
\begin{equation}
\label{hiperbola}
{m_\tau\over m_Z} + {m_\mu\over m_W}= 
{m_\tau\over m_\mu} a_\mu^{s.e.} + {m_\mu\over m_e} a_\mu^{v.p.}
\end{equation}
Note that $a_\mu^{v.p.}$, containing the vacuum polarisation (and light by
light) terms, has also an internal dependence on the quotients
$m_e/m_\mu$, $m_\tau/m_\mu$. 

{\bf Remark 3}
As the pure self-energy contributions do not depend (in QED) of lepton
mass, it is indifferent if we extract then from $a_e$ or $a_\mu$. Along this
note we have kept with $a_e$ due to historical reasons, but it results more
symmetric to refer to $a_\mu^{QED, s.e.}$ and $a_\mu^{QED, v.p.}$, as we
have done in formula (\ref{hiperbola}) above.

\end{document}